\documentclass[12pt]{article}
\usepackage{graphicx}

\def\beq{\begin{equation}}
\def\enq{\end{equation}}
\def\beqa{\begin{eqnarray}}
\def\enqa{\end{eqnarray}}

\def\MeV{\nobreak\,\mbox{MeV}}
\def\GeV{\nobreak\,\mbox{GeV}}

\def\mix{\lag\bar{q}g\si.Gq\rag}

\def\G3{\lag g^3G3\rag}

\def\pli{p^\prime}

\def\si{\sigma}

\def\al{\alpha}

\def\lb{\label}
\def\nn{\nonumber}

\newcommand{\rag}{\rangle}
\newcommand{\lag}{\langle}

\def\pbnr{}
\def\speaker{J. M. Dias $^{a}$, F. S. Navarra $^{a}$, M. Nielsen $^{a}$, C. Zanetti $^{b}$}
\def\onbehalfof{}
\def\title{$Z_{c}(3900)$ as a tetraquark state: decay width in \\QCD sum rules}
\def\affiliation{$^{a}$ Instituto de F\'isica, Universidade de S\~ao Paulo, C.P. 66318, \\
05389-970 S\~ao Paulo, S\~ao Paulo, Brazil\\
$^{b}$ Faculdade de Tecnologia, Universidade do Estado do Rio de 
Janeiro,\\
 Rod. Presidente Dutra Km 298, P\'olo Industrial, 27537-000,\\ 
Resende, Rio de Janeiro, Brazil}

\def\support{}

\newcommand\pubnumber{\pbnr}
\newcommand\pubdate{\today}
\def\Title#1{\begin{center} {\Large #1 } \end{center}}
\def\Author#1{\begin{center}{ \sc #1} \end{center}}

\newcommand{\OnBehalf}[1]{\sbox0{#1}\ifdim\wd0=0pt
        {}
	\else
	{\\on behalf of #1}
	\fi}
\newcommand{\SupportedBy}[1]{\sbox0{#1}\ifdim\wd0=0pt
        {}
	\else
	{\footnote{#1}}
	\fi}
\def\Address#1{\begin{center}{ \it #1} \end{center}}

\newcommand\pubblock{\includegraphics[width=5cm]{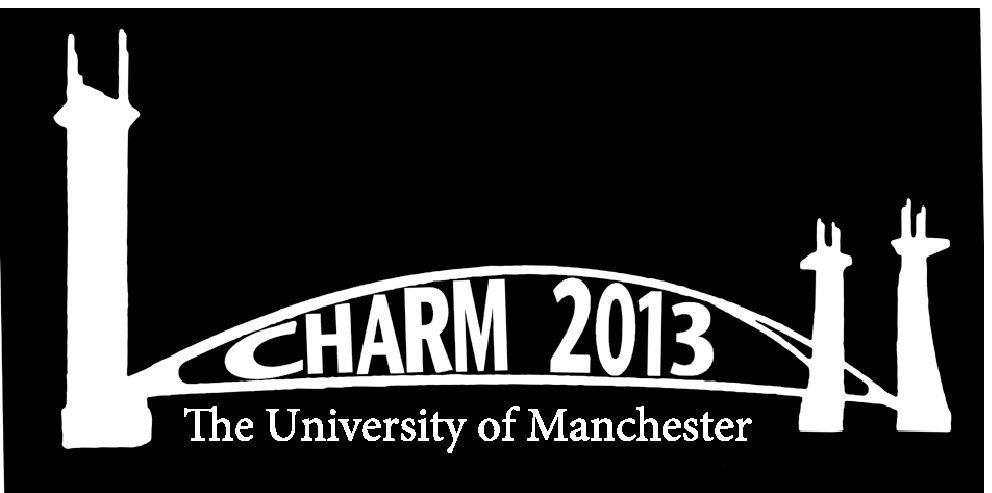}\hfill{\begin{tabular}{l} \pubnumber\\
         \pubdate  \end{tabular}}}
\newenvironment{Abstract}{\begin{quotation}  }{\end{quotation}}
\newenvironment{Presented}{\begin{quotation} \begin{center} 
             PRESENTED AT\end{center}\bigskip 
      \begin{center}\begin{large}}{\end{large}\end{center} \end{quotation}}
\def\Acknowledgements{\bigskip  \bigskip \begin{center} \begin{large}
             \bf ACKNOWLEDGEMENTS \end{large}\end{center}}
\def\venue{The 6$^{th}$ International Workshop on Charm Physics\\
(CHARM 2013)\\
Manchester, UK,  31 August -- 4 September, 2013}

\def\beq{\begin{equation}}
\def\eeq#1{\label{#1}\end{equation}}
\def\eeqn{\end{equation}}

\def\beqa{\begin{eqnarray}}
\def\eeqa#1{\label{#1}\end{eqnarray}}
\def\eeqan{\end{eqnarray}}

\let\bar=\overbar

\def\Dslash{\not{\hbox{\kern-4pt $D$}}}
\def\dslash{\not{\hbox{\kern-2pt $\del$}}}

\def\msb{{\bar{\ssstyle M \kern -1pt S}}}

\begin{document}
\begin{titlepage}
\pubblock

\vfill
\Title{\title}
\vfill
\Author{\speaker\SupportedBy{\support}\OnBehalf{\onbehalfof}}
\Address{\affiliation}
\vfill
\begin{Abstract}
Supposing the $Z_{c}^{+}(3900)$ as a charged partner of the X(3872), we use the QCD sum rules techinques in order to obtain the coupling constants of the $Z_{c}^{+} \, J/\psi\, \pi^{+}$, $Z_{c}^{+}\, \eta_{c}\, \rho^{+}$ and $Z_{c}^{+}\,D^{+}\, \bar{D}^{*}$ vertices and the corresponding decay width of these channels. We considered that the $Z_{c}(3900)$ can be described by a tetraquark current with the $J^{PC}=1^{+-}$ quantum numbers. As a result, we obtained a decay width value in good agreement with the experimental values reported by BESIII and Belle collaborations.

\end{Abstract}
\vfill
\begin{Presented}
\venue
\end{Presented}
\vfill
\end{titlepage}
\def\thefootnote{\fnsymbol{footnote}}
\setcounter{footnote}{0}
%

\section{Introduction}


	Very recently BESIII \cite{Ablikim:2013mio} and Belle \cite{Liu:2013dau} collaborations at the same time reported the observation of a charged state in the $Y(4260)\rightarrow J/\psi \pi^{+}\pi^{-}$ decay channel. This charged state was called $Z_{c}^{+}(3900)$. This observation was also confirmed by the authors of the Ref. \cite{Xiao:2013iha} using CLEO-c data. Hence, the $Z_{c}^{+}(3900)$ is the first charged state which observation was confirmed for more than one collaboration. 


The authors of the Ref. \cite{x3872} used the standard QCD sum rules technique in order to calculate the mass of a tetraquark structure with $J^{PC}=1^{+-}$ quantum numbers. As a result, they obtained a value in good agreement with the experimental mass value of the X(3872). They concluded that the X(3872) could be described by a tetraquark current. As the $Z_{c}$ has the same $J^{PC}$ quantum numbers as the X(3872), it can be considered as its charged partner. Therefore, the QCD sum rules result for the $Z_{c}$ mass is exactly the same result as in the X(3872) case. Therefore, to provide more evidence to support that the $Z_{c}$ can be explained as a tetraquark state, in this work we use the QCD sum rules to give an estimative about the $Z_{c}$ decay width. We will consider the following decay channels: $Z_{c}^{+} \, J/\psi \,\pi^{+}$, $Z_{c}^{+} \,\eta_{c} \,\rho^{+}$ and $Z_{c}^{+}\,D^{+}\, \bar{D}^{*}$. 


\section{Three-point function and $Z_{c}^{+}$ decay width}

In these three channels, there is always a vector and pseudoscalar mesons as final states. For the last two cases, the pseudoscalar mesons are described by the following pseudoscalars currents:
\begin{equation}
j_{5}^{\eta_{c}}=i\bar{c}_{a}\gamma_{5}c_{a}, ~  ~ j_{5}^{D}=i\bar{d}_{a}\gamma_{5}c_{a}.
\end{equation}
Regarding the pion, it is well established that the pion cannot be described, in QCD sum rules, by a pseudoscalar current. Therefore, in order to write the three-point function associated with the $Z_{c}^{+}  J/\psi \pi^{+}$ vertex, we use an axial current for describe the pion
\begin{equation}
j_{5\nu}^{\pi}=\bar{d}_{a}\gamma_{5}\gamma_{\nu}u_{a}.
\end{equation}
For the vector mesons, we use the following interpolating currents
\begin{equation}
j_{\mu}^{\psi}=\bar{c}_{a}\gamma_{\mu}c_{a}, ~ j_{\mu}^{\rho}=\bar{d}_{a}\gamma_{\mu}u_{a}, ~  j_{\mu}^{D^{*}}=\bar{c}_{a}\gamma_{\mu}u_{a}. 
\end{equation}
The interpolating current associated with the $Z_{c}^{+}(3900)$ considered in this work as a tetraquark structure, is given by
\begin{equation}
j_\alpha={i\epsilon_{abc}\epsilon_{dec}\over\sqrt{2}}[(u_a^TC\gamma_5c_b)
(\bar{d}_d\gamma_\alpha C\bar{c}_e^T)-(u_a^TC\gamma_\alpha c_b)
(\bar{d}_d\gamma_5C\bar{c}_e^T)]\;,
\label{field}
\end{equation}
where $a,~b,~c,~...$ are color indices, and $C$ is the charge conjugation matrix. 

In order to estimate the decay width of the channels mentioned earlier, we start writting the three-point correlation function associated for all the vertices
\begin{equation}
\Pi_{\mu\nu\alpha}(p,p^{'},q)=\int d^{4}x d^{4}y e^{ip^{'} \cdot x} e^{iq \cdot y } \Pi_{\mu\nu\alpha}(x,y),
\label{3p1}
\end{equation}
with
\begin{equation}
\Pi_{\mu\nu\alpha}(x,y)=\langle 0|T[j_{\mu}^{\psi}(x)j_{5\nu}^{\pi}(y)j_{\alpha}^{\dagger}(0)]|0\rangle,
\label{pxy}
\end{equation}
\begin{equation}
\Pi_{\mu\alpha}(x,y)=\langle 0|T[j_{5}^{\eta_{c}}(x)j_{\mu}^{\rho^{+}}(y)j_{\alpha}^{\dagger}(0)]|0\rangle,
\end{equation}
\begin{equation}
\Pi_{\mu\alpha}(x,y)=\langle 0|T[j_{5}^{D}(x)j_{\mu}^{D^{*}}(y)j_{\alpha}^{\dagger}(0)]|0\rangle.
\end{equation}
In Eq. (\ref{3p1}) $p=p^{'}+q$.

Once we have defined the three-point function for the vertices of interest, the next step is evaluate this function in two diferent ways: taking into account the hadronic degrees of freedom (phenomenological side) and also the QCD degrees of freedom. In the latter, we also use the Wilson's operator product expansion (OPE) to deal with the complex structure of QCD vacuum and for this reason it is frequently called OPE side. In order to illustrate how to use the QCD sum rules technique to extract coupling constants, we will show the details of the calculations to extract the coupling constant of the $Z_{c}^{+} J/\psi \pi^{+}$ vertex. For the other vertices, the procedure is analogous.

In order to obtain the phenomenological side, we insert a complete set of intermediate states for the $Z_{c}^{+} $, $J/\psi$ and $\pi^{+}$ into Eq. (\ref{3p1}). Using the following relations 
\beqa
\lag 0 | j_\mu^\psi|J/\psi(\pli)\rag =m_\psi f_{\psi}\varepsilon_\mu(\pli), ~ ~ \lag 0 | j_{5\nu}^\pi|\pi(q)\rag =iq_\nu F_\pi,\nn
\enqa 
\begin{equation}
\lag Z_c(p) | j_\alpha|0\rag =\lambda_{Z_c}\varepsilon_\al^*(p),\nonumber
\lb{fp}
\end{equation}
we get the following expression for the phenomenological side
\beqa
\Pi_{\mu\nu\al}^{(phen)} (p,\pli,q)={\lambda_{Z_c} m_{\psi}f_{\psi}F_{\pi}~
g_{Z_c\psi \pi}(q^2)q_\nu
\over(p^2-m_{Z_c}^2)({\pli}^2-m_{\psi}^2)(q^2-m_\pi^2)}
\nn\\
~\left(-\varepsilon_\mu(\pli)\varepsilon^*_\lambda(\pli)+{\pli_\mu \pli_\lambda
\over m_{\psi}^2}\right)\left(-\varepsilon^*_\alpha(p)\varepsilon^\lambda(p)
+{p_\alpha p^\lambda\over m_{Z_c}^2}\right)
+\cdots\;,
\label{phen}
\enqa
where the dots stand for the contributions for all excited states. The form factor, $g_{Z_c\psi \pi}(q^2)$, is defined by the generalization of the on-mass-shell matrix element, $\lag J/\psi \,  \pi \,| \, Z_c\rag$,
for an off-shell pion:
\beq
\lag J/\psi(\pli) \pi(q)|Z_c(p)\rag=g_{Z_c\psi \pi}(q^2)
\varepsilon^*_\lambda(\pli)\varepsilon^\lambda(p),
\label{coup}
\enq
where $\varepsilon_\alpha(p),~\varepsilon_\mu(\pli)$ are  the polarization
vectors of the $Z_c$ and $J/\psi$, respectively.

\begin{figure}[htb]
\centering
\includegraphics[height=1.5in]{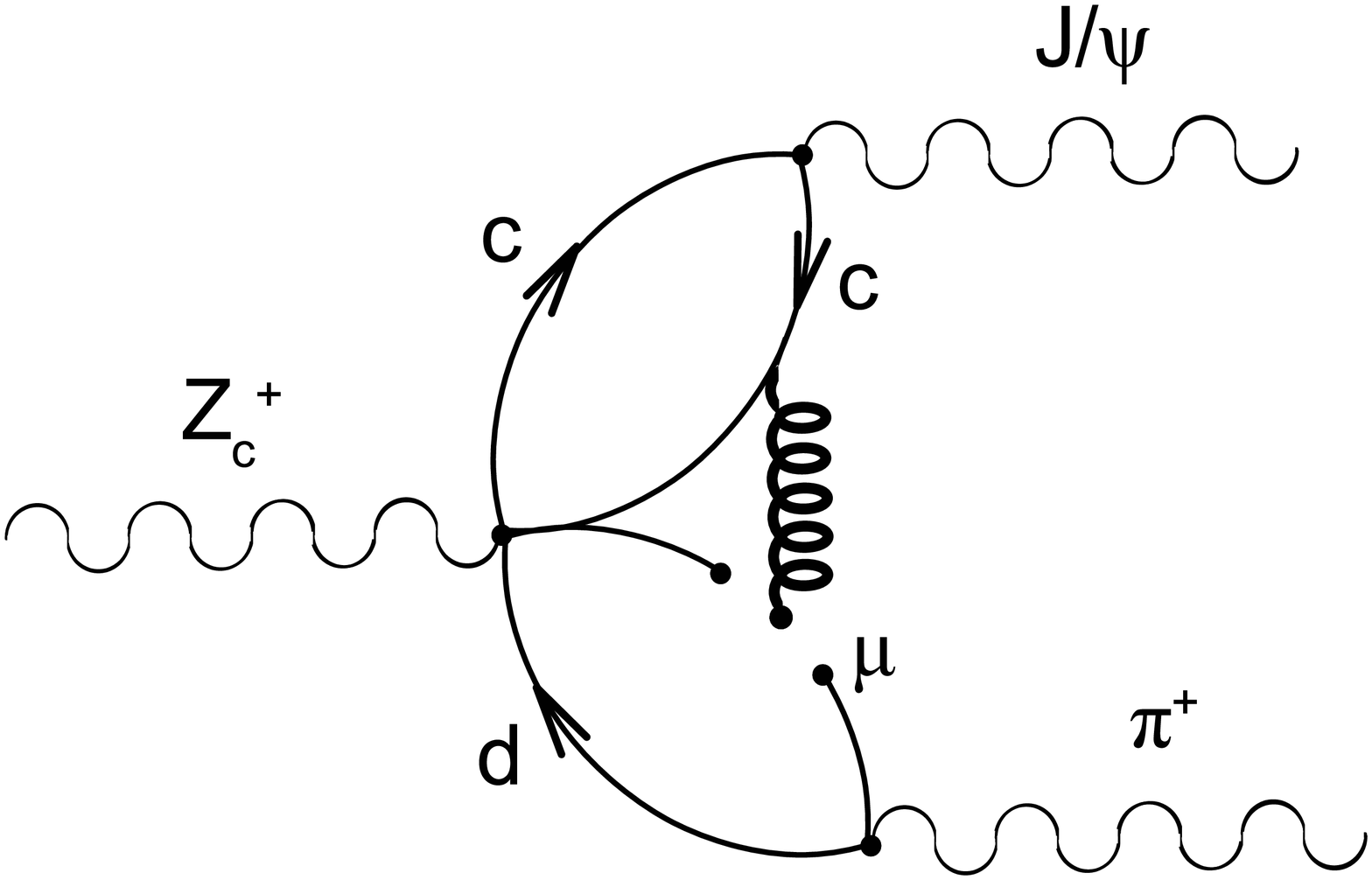}
\caption{Color-connected (CC) diagram which contributes for the OPE side of the sum rule.}
\label{fig:ccdiagram}
\end{figure}

In order to describe a genuine tetraquark in QCD sum rules,  in the OPE side we consider only the diagrams as the illustrated in Fig. \ref{fig:ccdiagram}. These type of diagrams are called color-connected (CC) diagrams. The reason is to maintain the nontrivial color structure, in QCD sum rules calculations, of the tetraquark current defined in Eq. (\ref{field}). In fact, due to the Fierz transformation, the tetraquark current can be rewritten in terms of a sum of molecular-type currents with a trivial color structure. 

The diagram in Fig. \ref{fig:ccdiagram} contributes only to the structures $q_\nu g_{\mu\alpha}$
and $q_\nu\pli_\mu\pli_\alpha$ appearing in the phenomenological side. We chose to work with
the $q_\nu\pli_\mu\pli_\alpha$ structure because structures with more momenta are supposed to work better. Therefore, the OPE side of the Eq. (\ref{3p1}) in the $q_\nu\pli_\mu\pli_\alpha$ structure is
\beq
\Pi^{(OPE)}= {\mix\over12\sqrt{2}\pi^2}{1\over q^2}
\int_0^1 d\alpha{\alpha(1-\al)\over m_c^2-\al(1-\al){\pli}^2}.
\label{ope}
\enq

If we neglect the pion mass in the right-hand side of Eq. (\ref{phen}), is possible to extract directly the coupling constant, $g_{Z_{c}J/\psi\pi}$, instead the form factor. Therefore, isolating the $q_\nu\pli_\mu\pli_\alpha$ structure in Eq. (\ref{phen}) and
making a single Borel transformation to both $P^2={P^\prime}^2\rightarrow M^2$,
we finally get the sum rule:
\beqa
A\left(e^{-m_\psi^2/M^2}-e^{-m_{Z_c}^2/M^2}\right)+B~e^{-s_0/M^2}=
{\mix\over12\sqrt{2}\pi^2}
\int_0^1 d\alpha \,  e^{- m_c^2\over al(1-\al)M^2},
\label{sr}
\enqa
where $s_0$ is the continuum threshold parameter for $Z_c$,
\beq
A={g_{Z_c\psi \pi}\lambda_{Z_c} f_{\psi}F_{\pi}~(m_{Z_c}^2+m_\psi^2)
\over 2m_{Z_c}^2m_{\psi}(m_{Z_c}^2-m_{\psi}^2)},
\label{a}
\enq
and $B$ is a parameter introduced to take into account single pole
contributions associated with pole-continuum transitions.


For the meson masses and decay constants we use the experimental values
\cite{pdg} $m_{\psi}=3.1$ GeV, $m_\pi=138$ MeV, $f_{\psi}=0.405$ GeV and
$F_\pi=131.52$ MeV. For the $Z_c$ mass we use the value measured in
\cite{Ablikim:2013mio}: $m_{Z_c}=(3899\pm6)$ MeV. The meson-current coupling,
$\lambda_{Z_c}$, defined in Eq. (\ref{fp}), can be determined from the two-point
sum rule \cite{x3872}: $\lambda_{Z_c}=(1.5\pm0.3)\times10^{-2}~\GeV^5$.
For the continuum threshold we use $s_0=(m_{Z_c}+\Delta s_0)^2$, with $\Delta s_0=
(0.5\pm0.1)~\GeV$.
\begin{figure}[htb]
\centering
\includegraphics[height=2.5in]{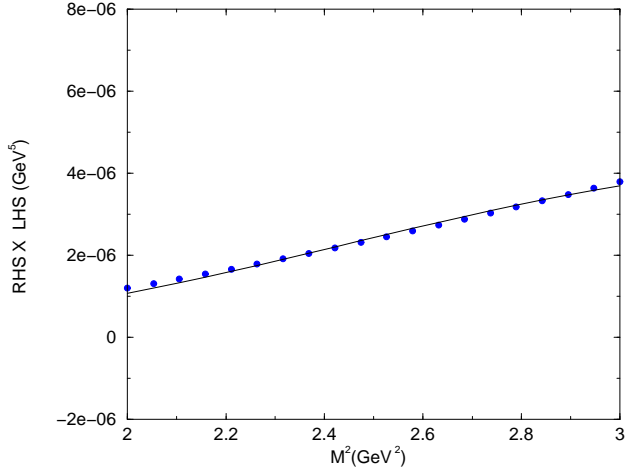}
\caption{Dots: the RHS of Eq.(\ref{sr}), as a function of the Borel mass
for $\Delta s_0=0.5$ GeV.
The solid line gives the fit of the QCDSR results through
the LHS of Eq.(\ref{sr}).}
\label{fig:fit}
\end{figure}


	To determine the coupling constant $g_{Z_c\psi \pi}$, we fit the QCD sum rules results
with the analytical expression in the left-hand side (LHS) of Eq. (\ref{sr}),
and find (using $\Delta s_0=0.5~\GeV$): $A=1.46\times10^{-4}~\GeV^5$ and
$B=-8.44\times10^{-4}~\GeV^5$.
Using the definition of $A$ in Eq. (\ref{a}), the value obtained for the
coupling constant is $g_{Z_c\psi \pi}=3.89~\GeV$, which is in excellent agreement
with the estimate made in \cite{maiani}, based on dimensional arguments.
Considering the uncertainties given above, we finally find:
\beq
g_{Z_c\psi \pi}=(3.89\pm0.56)~\GeV.
\label{coupling}
\enq

The decay width is given by:
\beqa
\Gamma(Z_c^+(3900)\to J/\psi\pi^+)={p^*(m_{Z_c},m_\psi,m_\pi)\over8\pi m_{Z_c}^2}
\times{1\over3}g^2_{Z_c\psi \pi}\left(
3+{(p^*(m_{Z_c},m_\psi,m_\pi))^2\over m_{\psi}^2}\right)\\
\enqa
where
\beq
p^*(a,b,c)={\sqrt{a^4+b^4+c^4-2a^2b^2-2a^2c^2-2b^2c^2}\over 2a}.
\enq
\vskip0.5cm
Therefore we obtain:
\beq
\Gamma(Z_c^+(3900)\to J/\psi\pi^+)=(29.1\pm8.2)~\MeV.
\label{width}
\enq

The results for the coupling constant and the decay width for the other channels are shown in Table \ref{tab:coupdecay}. Since the results for the vertex $Z_{c}^{+}(3900)\bar{D}^{0}D^{*+}$ it is exactly the same result as in the $Z_{c}^{+}(3900)D^{+}\bar{D}^{*0}$, we have also included it in this table. Therefore, the $Z_{c}$ total decay width is the sum of all partial decay widths which values are given in Table \ref{tab:coupdecay}. We get: $\Gamma_{Z_{c}}=(63 \pm 18.1)$ MeV.

\section{Summary and conclusions}

	We have used the QCD sum rules technique in order to give an estimative about the decay width of the very recently observed charged structure, $Z_{c}^{+}(3900)$. In particular, we evaluate the three-point function and extract the coupling constants of the $Z_{c}  J/\psi \pi^{+}$, $Z_{c} \eta_{c} \rho^{+}$ and $Z_{c}D^{+} \bar{D}^{*}$ and $Z_{c}^{+}(3900)\bar{D}^{0}D^{*+}$ vertices and the corresponding decay widths in these channels. As a result, the $Z_{c}^{+}$ total decay width is in good agreement with the experimental values reported by BESIII \cite{Ablikim:2013mio} and Belle \cite{Liu:2013dau} collaborations. 
\begin{table}[t]
\begin{center}
\begin{tabular}{l|ccc}  
vertex & coupling constant (GeV) &  decay width (MeV) \\ \hline
 $Z_{c}^{+}(3900)J/\psi \pi^{+}$  &   $3.89\pm 0.56$  &  $29.1 \pm 8.2$  \\
 $Z_{c}^{+}(3900)\eta_{c} \rho^{+}$ &   $4.85\pm 0.81$  &  $27.5 \pm 8.5$ \\
 $Z_{c}^{+}(3900)D^{+}\bar{D}^{*0}$ & $2.5\pm 0.3$  &  $3.2 \pm 0.7$ \\
 $Z_{c}^{+}(3900)\bar{D}^{0}D^{*+}$ & $2.5\pm 0.3$  &  $3.2 \pm 0.7$ \\  \hline
\end{tabular}
\caption{Coupling constant and decay width values for the corresponding channels.}
\label{tab:coupdecay}
\end{center}
\end{table}

\Acknowledgements
This work was supported by the Brazilian funding agencies FAPESP and CNPq.

\end{document}